# Infrasound of a Wind Turbine Reanalyzed as Power Spectrum and Power Spectral Density

**Comment on Pilger and Ceranna [1]:**
**The influence of periodic wind turbine noise on infrasound array measurements**
**(JSV, Vol. 388, pp. 188–200, 2017)**

Johannes Baumgart[a)], Christoph Fritzsche[b)], Steffen Marburg[c]

[a)] Saxon State Ministry for Energy, Climate Protection, Environment, and Agriculture, 01076 Dresden, Germany

[b)] Saxon State Agency for Environment, Agriculture, and Geology, 01311 Dresden, Germany

[c)] Chair of Vibroacoustics of Vehicles and Machines, Department of Mechanical Engineering, Technical University of Munich, 80333 Munich, Germany

**The infrasound levels due to the blade-tower interaction generated by a wind turbine in the publication by Pilger and Ceranna (JSV, Vol. 388, pp. 188–200, 2017) have to be corrected to be interpreted as sound pressure level. Also, the electrical power of the wind turbine should be corrected for the high wind case to 660 kW. We provide a reanalysis of the measured data with a power spectrum showing levels for the low-frequency signal of the wind turbine about 34 dB below the original work. All measured levels at a distance of 200 m from the wind turbine's infrasound signal are well below the hearing threshold.**

## 1. Introduction

The noise of wind turbines travels across property lines and is audible in the neighborhood. Despite comprehensive research on wind turbine noise [2–6], the debate about potential severe adverse health effects continues. Most of the wind turbine noise is in the audible frequency range. Our ability to sense moderate levels of low frequencies fades away below 20 Hz. However, the rotor blade passing the tower generates a characteristic pressure signal with dominant harmonic frequencies below 20 Hz, in the so-called infrasound range [7]. In principle, such an infrasound signal could be sensed if the levels are high enough. In recent years, the debate in Germany rose and had controversy on the level of such infrasound emitted by wind turbines. In our view, this debate is mainly rooted in a misinterpretation of the infrasound signal of a wind turbine measured by Pilger and Ceranna [1].

The measured pressure in the vicinity of a running wind turbine at ground level mainly consists of the random signal due to the wind and the turbine's periodic signal. Already at a wind speed of a few meters per second, the dynamic pressure is in the order of several Pascals. The unsteadiness of the wind causes a corresponding fluctuating pressure signal. The rotor blade of a wind turbine moves freely through the air until it approaches the tower. If the blade passes by the tower, the pressure field changes smoothly due to the interaction by the subsonic flow. This temporal change of the pressure field is measurable in the close neighborhood of a wind turbine. The rotation speed and number of blades set the periodicity.

## 2.    Results

### 2.1.    Inconsistency in published data in original work

Pilger and Ceranna [1] measured the outdoor pressure signal nearby a wind turbine and presented a spectrum with a short excerpt of the time series in their work. The amplitudes of their spectrum are the basis for their discussion and have been referenced by others. It is unclear how they scaled their spectrum and how the reported values of sound pressure level (SPL) should be interpreted.

We begin with an order of magnitude estimate under the assumption that the excerpt of the time series in their upper figure 4 is a representative excerpt for the spectrum shown below for the high wind conditions. Impulsive peaks repeat with a frequency of about 1.3 Hz, consistent with the 26 rounds per minute of the high wind conditions. A conservative estimate of the root-mean-square pressure based on the peak-to-peak value of their figure 4 and a sinusoidal function yields about 0.07 Pa. This corresponds to a sound pressure level of about 71 dB by using the reference pressure of $2 \times 10^{-5}$ Pa. The time-series signal was filtered with a 0.5 Hz high-pass filter. For frequencies above the filter frequency, the spectrum in the lower part of the figure has distinct peaks with amplitudes above 80 dB. The interpretation as a power spectrum – as the label and units of the ordinate might suggest – is not consistent with Parseval's Theorem. The levels of the harmonics are higher than the estimated sound pressure level. Pilger and Ceranna [1] do not supply sufficient details in their work to identify the scaling of the presented spectrum unambiguously. The inconsistency and the incomplete description of the data analysis have motivated us to reanalyze the publicly available time-series data[1] [8].

### 2.2.    Power spectral density and power spectrum

A Fourier transformation allows us to present time series data in the frequency domain. For the representation of the power as a function of frequency, a commonly used quantity in physics is the *power spectral density* (PSD). For random signals with peaks broader than the bandwidth, this quantity is continuous over the frequency and independent of the analysis bandwidth [9]. However, the PSD diverges at the peaks of discrete frequencies. For a sinusoidal signal, the PSD is a delta function. By integration over frequency, one obtains the *power spectrum* (PS), which is finite at the peak of a sinusoidal signal. However, for a random signal, the level of a PS decreases with the number of observed samples and is thus dependent on the windowing.

We have analyzed the data based on overlapped segmented averaging of modified periodograms [10] to obtain a spectrum with low variance. The long time series is analyzed by weighting segment by segment with a window function, performing a discrete Fourier transform, and averaging the squared pressure spectra. The discrete analysis has a finite bandwidth, depending on the sampling frequency, window length, and weighting function. Before the windowing, the subtraction of an average trend by linear regression removes the offset and possible leakage in adjacent frequency bands. For a discrete PSD the PS is obtained by integrating over each frequency bin separately. For simplicity, the amplitude is taken to be constant within the bin. Under this assumption, the integration results in power spectrum value times bin width. The spacing along the frequency axis, the bin width is set by the frequency resolution $f_{res}$, which is related to the sampling frequency $f_s$ and the window length $N$ by $f_{res} = f_s/N$. Finally, Parseval's theorem states that the power of a measured signal is equal to the sum of all components of the PS. If a window was applied, the components have to be divided by the normalized effective noise bandwidth [11] before the summation.

---

[1] To retrieve the data, for example, for the high wind condition:
http://eida.bgr.de/fdsnws/dataselect/1/query?station=HUF03&channel=HDF&starttime=2004-07-10T12:40:00&endtime=2004-07-10T13:10:00
with the corresponding metadata:
http://eida.bgr.de/fdsnws/station/1/query?station=HUF03&channel=HDF&starttime=2004-07-10T12:40:00&endtime=2004-07-10T13:10:00&level=response
The data can be converted to ascii by using mseed2ascii
https://github.com/iris-edu/mseed2ascii (accessed January 28, 2021).

For the reanalysis, the flat-top window HFT70 with a length of $N = 2^{14} = 16384$ samples and an overlap of 72.2 % was chosen, which has a high amplitude accuracy [10]. The original recording sampling frequency was $f_s = 100$ Hz [1] and was kept unchanged. This yields a bin width of $f_{res} = 0.0061$ Hz. In Fig. 1, the PS and PSD are plotted next to each other. We use $p_{ref} = 2 \times 10^{-5}$ Pa and the frequency of 1 Hz as references for the decibel scaling of PS and PSD. The standard definition of sound-pressure level as $SPL = 10 \cdot \log10(p_{RMS}^2/p_{ref}^2)$ and no frequency weighting was employed.

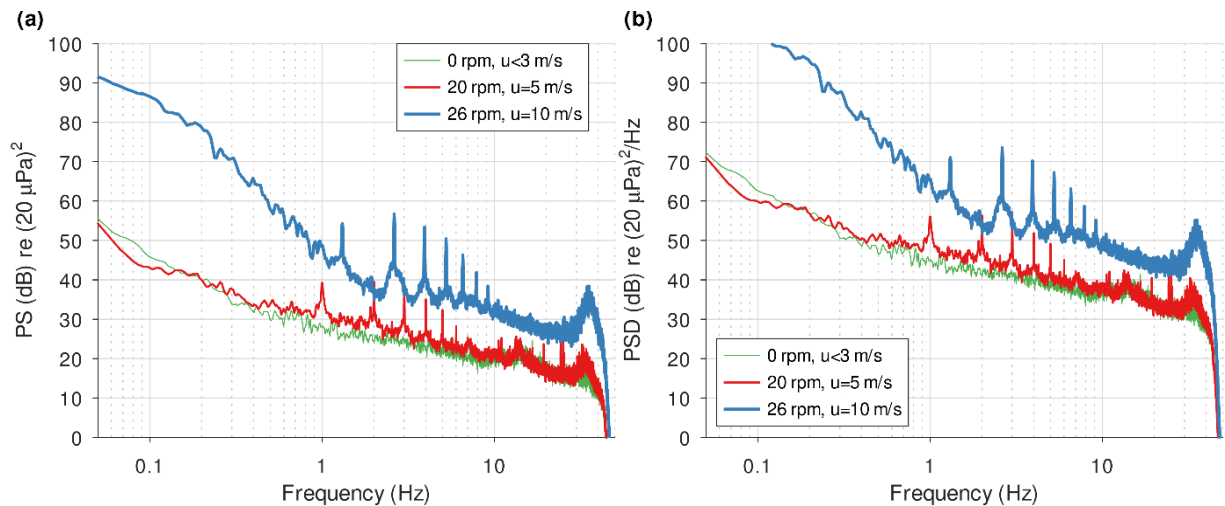

Figure 1: The power spectrum (PS, a) and the power spectral density (PSD, b) of pressure measurements recorded at a distance of about 200 m to a wind turbine. The PS and PSD differ only by a constant offset of 16.8 dB. The raw data originate from Pilger and Ceranna [1]. The turbine was turning with low rounds per minute (20 rpm, u = 5 m/s) or high (26 rpm, u = 10 m/s) or standing still (0 rpm, u < 3 m/s).

The noise bandwidth defines the constant scaling factor between PS and PSD [10], which is for this analysis 16.8 dB. This factor is set by the ratio $f_s/N$ and the normalized effective noise bandwidth of the used window [10], here 3.41. The spectra resemble in shape the original publication but have an offset of approximately -34 dB for the PS and -17 dB for the PSD with respect to the original publication.

## 2.3. Scaling of the spectra

A good agreement with the spectra presented by Pilger and Ceranna [1] in their figure 4 was achieved with a Hanning window with length $N = 8192$, an overlap of 50 %, and by showing the PSD times an estimated constant factor of about 50 by comparison with the original figure. Consistent with our analysis, this factor results in about 17 dB higher values than the PSD. The actual correction factor might differ by about 1 dB due to the graphical comparison.

Both spectra depend on the window length because the signal contains random and sinusoidal components. Fig. 2 depicts the first prominent peak region at about 1.3 Hz of the high wind spectrum.

The noise floor of the PS drops by about 3 dB for each doubling of the window length. The spacing on the frequency axis scales for the given sampling frequency inversely with the window length $N$. The width of the peak decreases with a longer window length while the amplitude is almost unaffected. On the other hand, the PSD provides an estimate of the noise floor, but the peak height depends on the window length. As the wind turbine generates a signal with small but finite variability, the peak has a finite width.

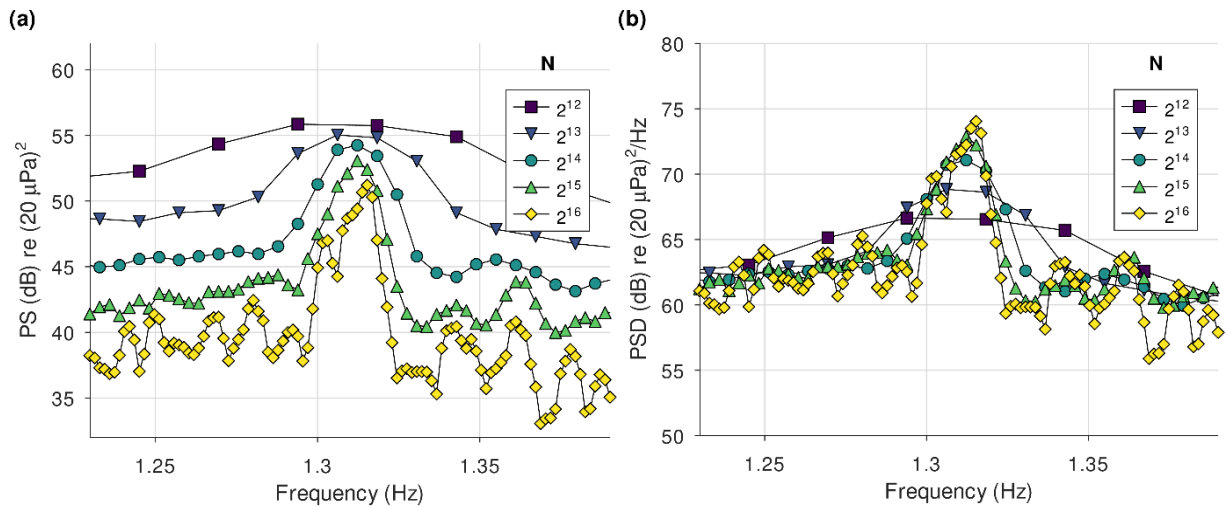

Figure 2: The power spectrum (PS, a) and the power spectral density (PSD, b) of Fig. 1 analyzed with different window lengths with 4096 to 65536 samples, covering the range from 0.0015 to 0.024 Hz for the bin width. The noise floor of the PS drops by about 3 dB for each doubling of the window length and is relatively unaffected at the peak. Contrary, the PSD provides an estimate of the noise floor, but the peak height depends on the window length. The raw data originate from Pilger and Ceranna [1].

## 3.  Discussion

### 3.1.  Amplitudes compared to hearing threshold

Today, we know more about how the pressure pulse due to the blade passing the tower is generated [7,12,13]. The key parameters are determined by the geometry of the blade and the tower in the configuration of the shortest distance. A common way to characterize a stationary signal is a spectrum of third octaves (Fig. 3). Published data of human hearing threshold at low frequencies [14] and the ISO 226 [15] provide estimates of the audibility of a signal at these low frequencies. Although the thresholds are generally based on sinusoidal tones, which differ to some degree from the measured signal, the levels are more than 20 dB below and, by this, clearly below the threshold's reference values. Above about 30 Hz the measured level is around the threshold, which is beyond the frequency range of the infrasound signal and possibly related to some other source than due to wind or the wind turbine's rotor blade passing the tower.

As a conservative upper bound, Fig. 3 provides the summing of the levels within the peaks of the harmonics additionally. For the high wind conditions, the sum around the fundamental and the seven distinct harmonics yields 63.1 dB, which is again well below the threshold of hearing in the corresponding frequency range. This is consistent with our conservative upper bound estimate of 71 dB for the short time series of the figure 4 of the original publication [1]. Walking causes similar pressure amplitudes by moving the head up and down by about a centimeter [16], as the pressure changes with height multiplied by density and gravitational constant. The equivalent upper bound estimate for the medium wind conditions of 46.3 dB is substantially lower.

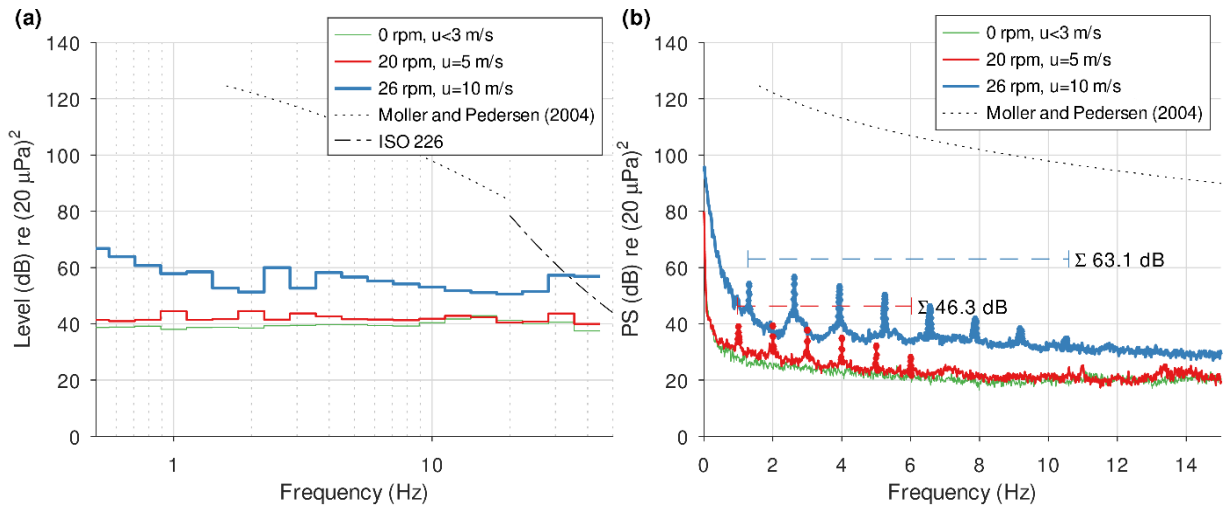

Figure 3: The sound pressure level analyzed in third octaves (a) and as power spectrum (b) in comparison to thresholds of hearing at low [14] and regular [15] frequencies. As integral value for the periodic signal, the levels in the vicinity of the peaks (marked by dots) are summed, and the integral value is indicated with the corresponding dashed line. The raw data originate from Pilger and Ceranna [1].

### 3.2. Electrical power of wind turbine

Additionally, we have reanalyzed the data at the neighboring stations using our approach and identified a similar decay with distance for the high wind conditions as in the work by Pilger and Ceranna [1]. This is in line with a constant offset independent of the measured signal.

Furthermore, it should be mentioned that the investigated wind turbine Vestas V47 has two electrical generators. One with 200 kW electrical power, which operates if the turbine runs at 20 rpm. In the mode with 26 rpm, a generator with 660 kW electrical power is in use. In the work by Pilger and Ceranna [1] only a value of 200 kW is reported. We suspect that the extrapolated values for higher electrical power in their figures 7 and 8 should be divided by the factor 3.3, which corresponds to subtracting 5.2 dB as an additional correction.

## 4. Conclusion

To sum up, the infrasound levels due to the blade-tower interaction generated by a wind turbine in the publication by Pilger and Ceranna [1] have to be corrected to be interpreted as sound pressure level. Also, the electrical power of the wind turbine should be corrected for the high wind case to 660 kW. We provide a reanalysis of the measured data with a power spectrum showing levels for the low-frequency signal of the wind turbine about 34 dB below the original work. All measured levels at a distance of 200 m from the wind turbine's infrasound signal are well below the hearing threshold.

## 5. Acknowledgment

We thank Kristy Hansen and all others who contributed by constructive comments on the manuscript and by critical checking the data analysis. This research did not receive any specific grant from funding agencies in the public, commercial, or not-for-profit sectors.